\documentstyle[12pt]{article}
\begin{document}
\def\beq{\begin{equation}}
\def\eeq{\end{equation}}

\begin{flushright}
ITEP-TH-56/99\\
hep-th/xxnnmmm
\end{flushright}
\vspace{0.5cm}

\begin{center}
{\Large\bf On heavy states in supersymmetric  gluodynamics at large N}

\bigskip{{\Large A.Gorsky and K.Selivanov } \\
ITEP, Moscow, 117259, B.Cheremushkinskaya, 25}
\end{center}
\bigskip

\begin{abstract}
It is argued that there are states (quasiparticles)
with masses ranging over
the scales $\Lambda N_{c}^{1/3} \div \Lambda N_{c}$ in N=1
supersymmetric multicolor gluodynamics. These states
exist in the form of quantum bubbles made out of the BPS domain walls.
Analogous states are likely to exist in non-supersymmetric case as well.
\end{abstract}

Recently a remarkable progress in understanding supersymmetric
gauge theories took place. One of the main roles in those
developments was played by the so-called BPS states which
showed up practically in every instance when some kind of exact information
about a gauge theory was available. The classical example was,
of course, the construction of low energy effective action in
N=2 supersymmetric Yang-Mills \cite{SW}, where a crucial piece of
information was obtained due to existence of the BPS saturated  monopoles
and dyons.

N=1 supersymmetric Yang-Mills also possesses the BPS states.
These are domain walls \cite{DS} and (upon perturbing the theory
with some operators or upon introducing some matter)
strings \cite{GS}.
Notice that this time the BPS states are not point-like which is,
of course, much harder to work with. It may easily happen that
extendedness of the BPS states in N=1 supersymmetric gluodynamics
is the reason why it is less treatable compared to the N=2 theory.
Anyway, the role of the BPS extended objects in dynamics
of N=1 theory is still to be understood, though there are
permanent advances (see, e.g. \cite{GaS}-\cite{KKS}) of which the most
inspiring for the present note was the paper \cite{GaS}).

The paper \cite{GaS}
was concentrated about the fact that the BPS domain wall
width scales as $1/N_{c}$ at large $N_{c}$. The point of the paper was
that such behavior of the width assumes existence of heavy particles,
$m \sim N_{c}$ (of which the wall could be done) in the spectrum
of multicolor N=1 supersymmetric gluodynamics, in addition
to the traditional glueballs whose mass is independent of $N_{c}$
in the multicolor limit. It is worth stressing that these heavy
particles were argued in  \cite{GaS}
to exist in the non-supersymmetric gluodynamics too.

The purpose of the present note is to suggest a candidate for
these heavy particles. We argue that they can, in turn, be
done out of the BPS walls, namely, in the form of quantum bubbles done
out of the BPS walls. It happens that at large $N_{c}$
these bubbles can be treated semiclassically, the thing
wall approximation being applicable (because a typical
radius of the bubble is much bigger than the wall width).
Masses of these bubbles range over the scales
$\Lambda N_{c}^{1/3} \div \Lambda N_{c}$.

The BPS domain walls in N=1 supersymmetric gluodynamics were introduced
in \cite{DS}. They interpolate between the $N_{c}$
vacua distinguished by the value of gluino condensate. The tension of the
wall between two adjacent vacua
is of the order of $N_{c}{\Lambda}^{3}$, where $\Lambda$ stands for the
scale of the theory. It is instructive to discuss the domain walls
at large $N_{c}$ in the framework of M-theory, where they are represented
as M5 branes wrapped on some cycles \cite{W}. This construction naturally
explains that the wall tension scales as $N_{c}$. Later
it was argued in refs. \cite{DGK} that the domain wall width
$\delta$
scales as $\delta \sim 1/N_{c}$. The arguments of  refs. \cite{DGK},
were supplemented in \cite{GaS} by considering the $N_{c}$-behavior \cite{GS}
of the wall junctions \cite{WJ} and by studying various effective
Lagrangians.

Quantum bubbles (in a scalar theory with spontaneous symmetry
breaking) were introduced in \cite{GV}, where they arose as
resonant states in the multi-particle production at a threshold.
Let us remind main points concerning those bubbles. The action in the
thin wall approximation  reads
\begin{equation}
\label{action}
S=4{\pi}\mu \int dtr^{2}\sqrt{1-{\dot{r}}^{2}}
\end{equation}
where $\mu$ stands for the tension of the wall.
Corresponding Hamiltonian reads
\beq
H^2-p^2=( 4{\pi} \, \mu \, r^{2})^2
\label{clm}
\eeq
with the canonical momentum $p$. The classical trajectory
with energy $E$ corresponds to oscillations of the bubble between the
turning point \newpage
$r=r_{0}=(\frac{E}{4{\pi}\mu})^{1/2}$  and $r=0$.
It could be that instead of oscillating the bubble would quickly
dissipate into outgoing waves. Below we give some arguments that
this is not the case.
The bubble becomes stable in the large $N_{c}$ limit.
Therefore for the moment we proceed discussing bubbles
in assumption of their stability.

The part of the trajectory near
zero radius, $r \sim \delta$, cannot be described within the thin wall
approximation. However in the case that
$r_{0} \gg \delta$ (which we show below is indeed the case) most
of the evolution of the bubble proceeds within the applicability of the thin
wall approximation. Via standard quasi-classical consideration this
assumes that the wave function of the bubble decays quickly near zero
radius so this region should not be essential.

The oscillatory motion of the bubbles can be quantized and the discrete
energy levels found by applying the Bohr - Sommerfield quantization rule:

\beq
I(E) \equiv \oint p \, dr - 2\pi \, \nu(E)= 2\pi \, n~,
\label{bs}
\eeq
where the integral runs over one full period of oscillation and contains the
momentum $p$ determined by eq.(\ref{clm}) in the thin wall approximation.
The quantity $\nu(E)$ is a correction to the thin wall limit, which arises
from the contribution to the action of the motion at short distances $r \sim
\delta$, where the latter limit is not applicable. Since at such
distances $p \sim E$, by order of magnitude $\nu(E)$ can be estimated as
$\nu(E) \sim E \delta$. The integral in eq.(\ref{bs}) is of the order of
$E\,r_0$, and is thus much larger than $\nu(E)$ once the condition $r_0 \gg
\delta$ is satisfied. In terms of the turning radius $r_0$ the quantization
relation (\ref{bs}) reads as

\beq
k \mu r_0^{3} = 2 \pi \, (n+\nu(E))~,
\label{bsr}
\eeq
with $k$ being a numerical coefficient,

\beq
k={{\pi}{\sqrt{\pi}\, \Gamma[1/4]} \over {\Gamma[7/4]}}~.
\label{kd}
\eeq
For an energy level $E_n$ in terms of the number $n$ of the level one finds:
\beq
E_n=\mu^{1/3}\, \left ( {{2 \pi \,n}/{k}} \right )^{2/3}~.
\label{bse}
\eeq
Thus for finite $n$ the energy of the bubble scales as $N_{c}^{1/3}$.

Let us first verify that the thin wall approximation is valid.
Using the fact that the tension of the wall scales
as $N_{c}$, one sees from Eq.(\ref{bsr}) that (at finite $n$)
\beq
\label{radius}
r_{0} \sim N_{c}^{-1/3}
\eeq
which is indeed large compared to the wall width, $\delta \sim 1/N_{c}$
(on the other side, $r_{0}$ is small compared to the scale of the
theory, $\Lambda$, so the bubbles are point-like).

Let us now estimate the decay rate of the quantum bubble
into the light glueballs (whose mass is independent
of $N_{c}$ in the multicolor limit). The spectrum of the particles produced
by an accelerating wall is described by the exponential
$e^{-E/T_{eff}}$, where $E$ is the energy of the produced particles and
$T_{eff}$ stands for an effective temperature, which is equal to
the acceleration of the wall. In the case of the bubble a typical acceleration
is of the order of $1/r_{0}$. Hence, in view of Eq.(\ref{radius}),
we expect that the particle production rate is $e^{-CN_{c}^{1/3}}$ where
$C$ is $N_{c}$-independent.

It is seen from Eq.(\ref{bsr}) that for sufficiently high levels of the
bubble, $n \sim N_{c}$, $r_{0}$ becomes unsuppressed by the powers
of $N_{c}$ and hence for such levels one expects no suppression
of the particle production. Remarkably, energy of these
extremal levels scales as $N_{c}$, which is the mass scale
argued  in \cite{GaS} to exist in N=1 supersymmetric multicolor
gluodynamics in order to explain the $N_{c}$ dependence of
the domain wall width.

It is worth noticing that both perturbative and nonperturbative
corrections to the effective action Eq.(\ref{action}) are likely to be small.
Indeed, the effective coupling constant is expected to freeze at the scale
$r_{0}$ which is small compared to $1/ \Lambda$.
So, perturbative corrections are expected to be of the order of
$1/logN_{c}$. Nonperturbative corrections should be suppressed
by powers of $N_{c}$.
One is tempted to
claim that in the multicolor limit Eq.(\ref{bse}) is nonexact only
because it is obtained via Bohr-Sommerfield quantization, which is good
only for sufficiently high levels.
Notice, however, that Eq.(\ref{bse}) gives only part of the spectrum,
nonspherical modes of the bubble being neglected. The problem of
quantization of the bubble in general case is the problem of quantization of
supermembrane which is unsolved by now.

All of the above picture is very likely to exist in
the usual non-supersymmetric gluodynamics in the multicolor limit.
In \cite{W2} (see also \cite{S}, \cite{G}) it was argued that
even in the non-supersymmetric case at every $\theta$ where are $N_{c}$
vacua one of which being stable and others - metastable.
However those metastable vacua live infinitely long in the
multicolor limit. The wall interpolating between the adjacent vacua
was argued to have the same $N_{c}$-dependences -
$N_{c}$ in its tension and $1/N_{c}$ in its width.
Of course, the walls  between
nondegenerate vacua cannot be at rest, but the quantum bubbles can
still exist.
A typical size of the quantum bubbles is seen to be much less than the
size of the critical bubble (the one which arise in the spontaneous
metastable vacuum decay), hence the above estimates remain
intact.

Let us briefly discuss possible interactions of quantum
bubbles in Minkowski space. First we have to pose question
about the bubble charges. It is clear that flat domain
wall is charged with respect to three-form field which
can be identified in non-supersymmetric case with Chern-Simons
three-form. However the total charge of the bubble vanishes.
The most simple argument concerning this point amounts from
the analogy with the behaviour of
one-form field in two dimensions.
Indeed in d=2 electric field of the charge is constant and
the total field of charge-anticharge pair
is zero. The bubble
above is the analogue of this pair in d=4
and taking into account
that the curvature of the three-form field
of the  wall is constant
one can show the absence of the total three-form charge.

Therefore there is no the Coulomb like
interaction between two bubbles.
However one can look for the
possible string like interaction
for the bubble pair. Naively one
can expect a string stretched
between bubbles since QCD string can
end on the domain wall.
But more careful inspection shows that
it is not the case. The
point is that U(1) field providing
this possibility is related
to the presence of the fermionic zero
mode on the flat domain wall.
In the bubble case zero   mode
disappears due to the curvature
therefore the argument above
fails in this case. Thus there
is no interaction of the bubble
pair via a single QCD string.
On more supporting argument
comes from the consideration
of the sizes of the objects - the size of QCD string
is believed to be
independent of $N_{C}$  while
the size of the low level
bubble vanishes in the multicolor limit. The issue of other
possibilities for the bubbles to interact deserves further
investigation.

It would be interesting to realize the influence of these
additional states on the
thermodynamics of the multicolor
QCD. The fractional dependence on N implies the possibility
of a kind of the phase transition related with the quantum
bubbles. Since we deal with the large N
limit one can also expect
gravitational counterpart of these states via AdS/CFT
correspondence.

We would like to thank M. Shifman
and M. Voloshin for discussions
on the related issues.
The work of A.G. was supported in part by grant INTAS-97-0103 and
the work of K.S. by grant INTAS-96-0482.

\end{document}